\documentclass[aps,pra,twocolumn,showpacs]{revtex4}  % for review and submission
\usepackage{graphicx}  % needed for figures
\usepackage{dcolumn}   % needed for some tables
\usepackage{bm}        % for math
\usepackage{amssymb}   % for math
\usepackage{amsmath}
\usepackage{epstopdf}
\usepackage{color}
\usepackage{threeparttable}
\usepackage[sort&compress]{natbib}
\begin{document}

% The following information is for internal review, please remove them for submission
%\widetext
%\leftline{Version xx as of \today}
%\leftline{Primary authors: Joe E. Physics}
%\leftline{To be submitted to (PRL, PRD-RC, PRD, PLB; choose one.)}
%\leftline{Comment to {\tt d0-run2eb-nnn@fnal.gov} by xxx, yyy}
%\centerline{\em D\O\ INTERNAL DOCUMENT -- NOT FOR PUBLIC DISTRIBUTION}

% the following line is for submission, including submission to the arXiv!!
%\hspace{5.2in} \mbox{Fermilab-Pub-04/xxx-E}

\title{Non-destructive light shift measurements of single atoms in optical dipole traps}
%\input author_list.tex       % D0 authors (remove the first 3 lines
                             % of this file prior to submission, they
                             % contain a time stamp for the authorlist)
                             % (includes institutions and visitors)
\author{Chung-Yu Shih}
\author{Michael S. Chapman}
\affiliation{School of Physics, Georgia Institute of Technology, Atlanta, Georgia 30332-0430}
\vskip 0.25cm

\date{\today}

\begin{abstract}

We measure the AC-Stark shifts of the $5S_{1/2}, F=2 \rightarrow 5P_{3/2}, F'=3$  transitions of individual optically trapped ${}^{87}$Rb atoms using a non-destructive detection technique that allows us to measure the fluorescent signal of one-and-the-same atom for over 60 seconds. These measurements allow efficient and rapid characterization of single atom traps that is required for many coherent quantum information protocols. Although this method is demonstrated using a single atom trap, the concept is readily extended to resolvable atomic arrays.

\end{abstract}

\maketitle

\newcommand{\bra}[1]{\langle #1|}
\newcommand{\ket}[1]{|#1\rangle}
\newcommand{\braket}[2]{\langle #1|#2\rangle}

Optical dipole trapping of individual neutral atoms is an active area of research, motivated in large part by applications in quantum information science, quantum many-body physics, and investigations of foundational issues in quantum mechanics. Building upon the early demonstrations of single neutral atom traps \citep{Kimble1994,Meschede2000}, there have been many impressive advances with resolved atom traps including deterministic loading \citep{Schlosser2001}, manipulation and control of single atoms \citep{Meschede2012}, cooling \citep{Kaufman,Thompson}, non-destructive state measurement \citep{Gibbons2011,Browaeys2011}, multi-dimensional atomic registers \citep{Meschede2004,Nelson,Karski,Bakr}, cavity QED with individual trapped atoms \citep{Ye1999,Rempe2000,Chapman2007} and demonstration of atom-photon \citep{Weinfurter2006,Wilk2007} and atom-atom entanglement \citep{Urban,Gaetan,Wilk2010,Isenhower2010}.
Resolved atom optical dipole traps can be used to hold cold atoms for times exceeding 300~s \citep{Gibbons2008} and provide a promising alternative to rf trapped ion systems. However, one important difference compared with rf ion traps is that the optical trapping fields intrinsically shift the energy levels of the atoms (so-called `light shifts') and thereby alter both the energy levels in which the information is stored (typically ground state hyperfine levels) as well as the frequencies of the optical transitions needed to manipulate the atoms. Successful realization of large-scale neutral atom quantum registers will require measurements of these shifts for each individual atom in order to characterize the trapping environments. A variety of destructive techniques (in the sense that the trapped atom is lost) have been used to measure excited state AC-Stark shifts of trapped neutral atoms including trap ejection excitation techniques \citep{Kim2002}, absorption spectra of singly trapped ${}^{87}$Rb atoms \citep{Tey2008} and ionization from Rydberg states \cite{Younge2010,Markert2010}. For practical purposes, it will be necessary to employ non-destructive techniques for arrays of more than a few atoms.

The focus of this work is the development of an efficient technique to characterize the transition shifts for the typical case in which the differential shifts of the excited states are comparable to both the intrinsic linewidth of the transition and the shifts of the ground state. The principle novelty of our method is that it is non-destructive in the sense that a single atom can be used to obtain the fluorescent spectrum of entire excited state manifold. It employs a continuous stream of short alternating probe and cooling pulses (durations 1~$\mu$s and 99 $\mu$s respectively) together with gated single-photon detection that provides high signal to noise ratio ($>$20) and an atom lifetime $>$60 s even for resonant probe detunings. Using this method, we measure the AC-Stark shifts of the $F=2 \rightarrow F'=3$ D2 transitions in ${}^{87}$Rb for different trap depths and probe polarizations. The spectra are compared to theoretical calculations using independent measurements of the trap depth and atom temperature.

\begin{figure}
\includegraphics[width=85mm]{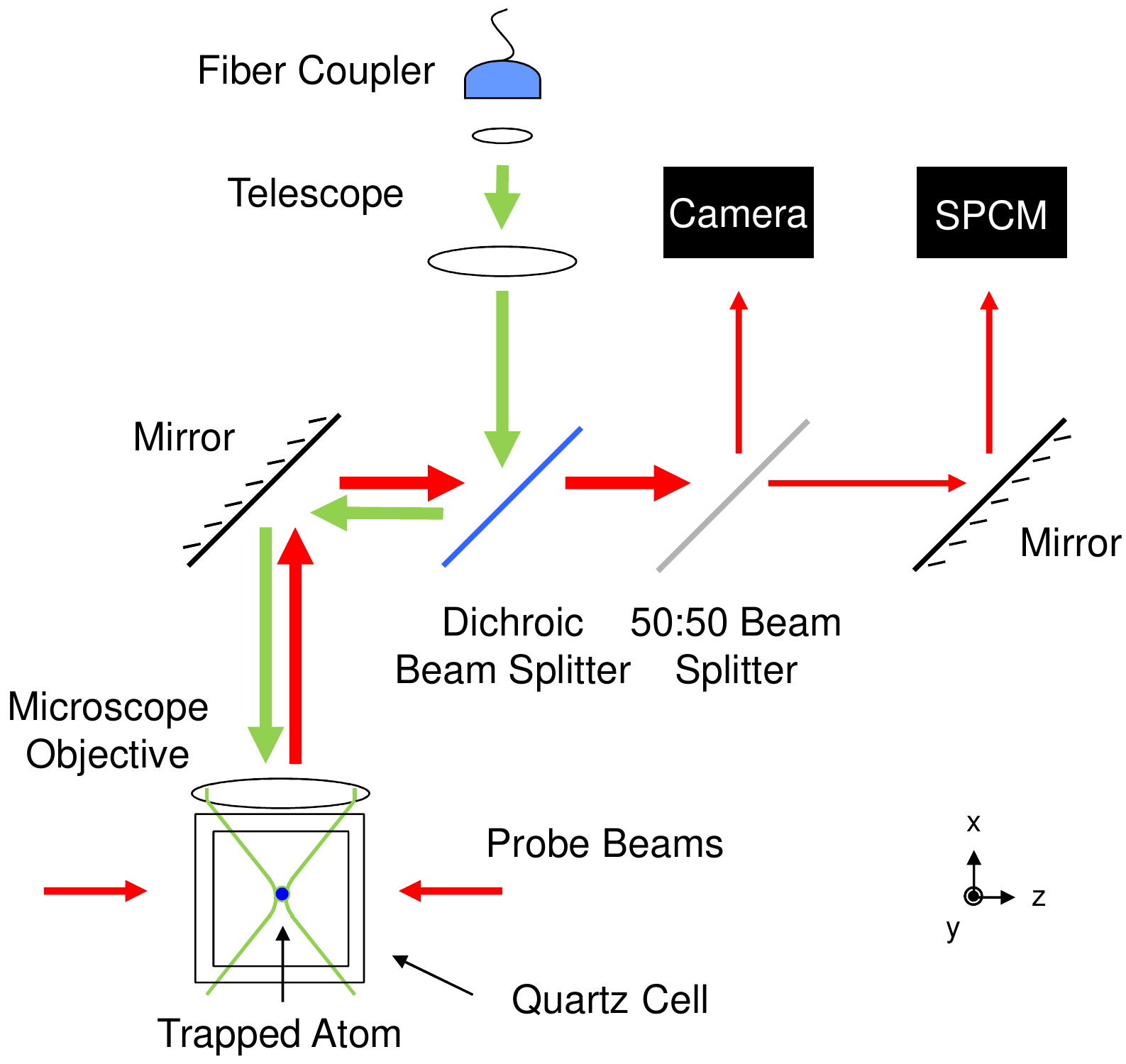}
\caption{\label{schematic} (Color online) The microscope objective is mounted outside the quartz cell. The 780 nm (red) fluorescent signal and 1064 nm (green) beam paths are separated with a dichroic beam splitter. The fluorescent signal is sent to a 50:50 beam splitter; half of the light goes to a EMCCD camera and the other half goes to the single photon counting module (SPCM).}
\end{figure}

\begin{figure}
\centering
\includegraphics[width=85mm]{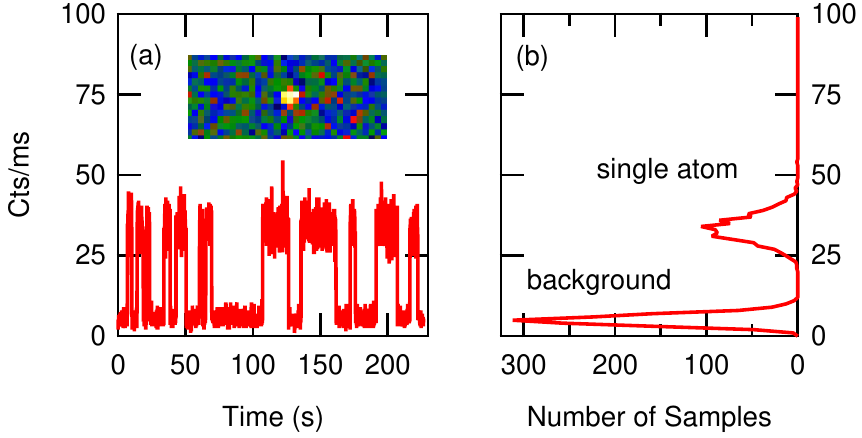}
\caption{\label{1atomintrap} (Color online) (a) The fluorescent signal of the atoms in the single focus trap when the FORT is continuously being loaded for over 200~s. The probe laser is 63.2 MHz blue detuned from the bare resonance and the probe intensity is $\simeq$2$I_{sat}$. The inset is an image of an atom in this trap taken with 1~s exposure time. (b) Histogram of the signal. The light-assisted two-body collision ensures only one atom is loaded into the trap.}
\end{figure}

The schematic of the experiment is illustrated in Fig.~\ref{schematic}. The magneto-optical trap (MOT) has a six-beam configuration, one pair of the optical molasses beams are along the $z$-axis (magnetic field gradient direction) and the other two pairs are in the $x$-$y$ plane. Each molasses beam has a intensity of 2~mW/cm$^{2}$ and a beam diameter of 1~mm. To create a single atom MOT, a field gradient of 250 G/cm is used, while a field gradient of 80 G/cm provides a well-localized MOT of $\sim$30 atoms localized to $<$50 $\mu$m.
A single focus far-off resonant trap (FORT) is created by focusing a $\lambda=$ 1064 nm fiber laser beam with a high numerical aperture microscope objective (NA = 0.4); the same objective is also used to collect the fluorescent signal from the trapped atoms. Cold samples of ${}^{87}$Rb atoms are initially loaded in the MOT using a MOT beam detuning of $-11$ MHz from the $F = 2$ $\rightarrow$ $F^{\prime} = 3$ cycling transition. Following loading of the MOT, the atoms are transferred to the FORT by turning off the magnetic field gradient and further red-detuning the MOT beams to $-23$ MHz for optimal cooling. The atoms are detected using both an EMCCD camera (Andor iXon) and a single photon counting module (SPCM). The total photon collection efficiency of the detection system is 0.9\%; half of the collected fluorescence is imaged on the camera and the other half is focused onto the SPCM. The atom in the single focus trap is localized to region of 2 by 2 pixels on the camera, which corresponds to 2.5 $\mu$m by 2.5 $\mu$m in the trapping region (see Fig.~\ref{1atomintrap}). The background scattering from the MOT beams has a average value of $10^4$ cts/s for the region of interest with fluctuations of 100 cts/s. The corresponding signal rate for a single atom in the FORT excited by the MOT beams is $\simeq$1000 cts/s on the camera, which yields a signal to noise ratio of 10 with 1~s exposure time.The tightly focused single focus trap operates in the collisional blockade regime which ensures only one atom is loaded each time \citep{Schlosser2001,Schlosser2002}. Fig.~\ref{1atomintrap} shows the loading dynamics of the single focus trap and histogram of the fluorescent signal obtained using the technique described in the next Section. The probe laser is $-3$ MHz detuned from the shifted resonance at $\sim+$66 MHz so that the photons scattered by atoms not in the FORT is negligible. It is evident that there is at most a single atom loaded into the single focus trap.
The trapping potential is characterized by measuring the radial and longitudinal trap frequencies \citep{Friebel1998} using parametric excitation \citep{Wu2006}. For a single focus trap with 70 mW trap laser power, the measured trap frequencies are ($\nu_r$, $\nu_z$) = (42.5, 3.5) kHz, which infers a $\simeq$2.5 $\mu$m minimum beam waist and $k_B\times$0.88 mK trapping potential (equivalent to $h\times18$ MHz).

Although single atoms can be easily resolved in fluorescent images with the camera, this requires long exposures and hence is limited to a small range of cooling beam detunings and beam geometries that provide continuous cooling of the atoms. Ideally, we would prefer a technique that would allow detection of single atoms excited with arbitrary detuning with a high repetition rate while keeping the same atom. In order to achieve this, a gated probing/cooling technique is employed using a dedicated pair of counter-propagating probe beams together with time-resolved fluorescent detection using the SPCM. The probe beams are aligned along the $z$-axis to minimize background scattering into the detection system and have a minimum waist of 125 $\mu$m. Using these probe beams, the maximum signal from a single atom is $\sim$40 cts/ms on the SPCM. For comparison, the scattering of the MOT beams off the quartz cell walls is $\sim$1000 cts/ms on the SPCM (with fluctuations of $\sim$30 cts/ms), which gives a signal to noise ratio of $\sim$1 for a 1~ms acquisition time. Although the signal to noise ratio could be improved by reducing the field of view of the SPCM (currently 50 $\mu$m by 50 $\mu$m), this comes at the price of greatly increasing the alignment difficulty of the imaging system. Instead, the MOT and the probe beams are switched on and off in an alternating manner, and the SPCM is gated on only during the probing period. After each probing period, the atoms are sub-Doppler cooled by the MOT beams to the bottom of the trap to suppress the probability of losing atoms due to heating from the probe excitation. The probe cooling cycle is 100 $\mu$s, during which the atom is probed for 1~$\mu$s and cooled for 99 $\mu$s. This reduces the total background scattering to less than 3~cts/ms, typically. The resulting signal to noise ratio is $\sim$20 for a 1~s acquisition time near the shifted resonance (limited by the photon shot noise of the signal), which is a factor of 20 improvement. This technique enables us to operate with a repetition rate up to 10 kHz while retaining a high signal to noise ratio. Furthermore, it allows us to probe single atoms over a large detuning range with a lifetime over 60 s, which is well-suited for measuring the light shifted spectrum of the optically trapped atoms.

\begin{figure}
\includegraphics[width=85mm]{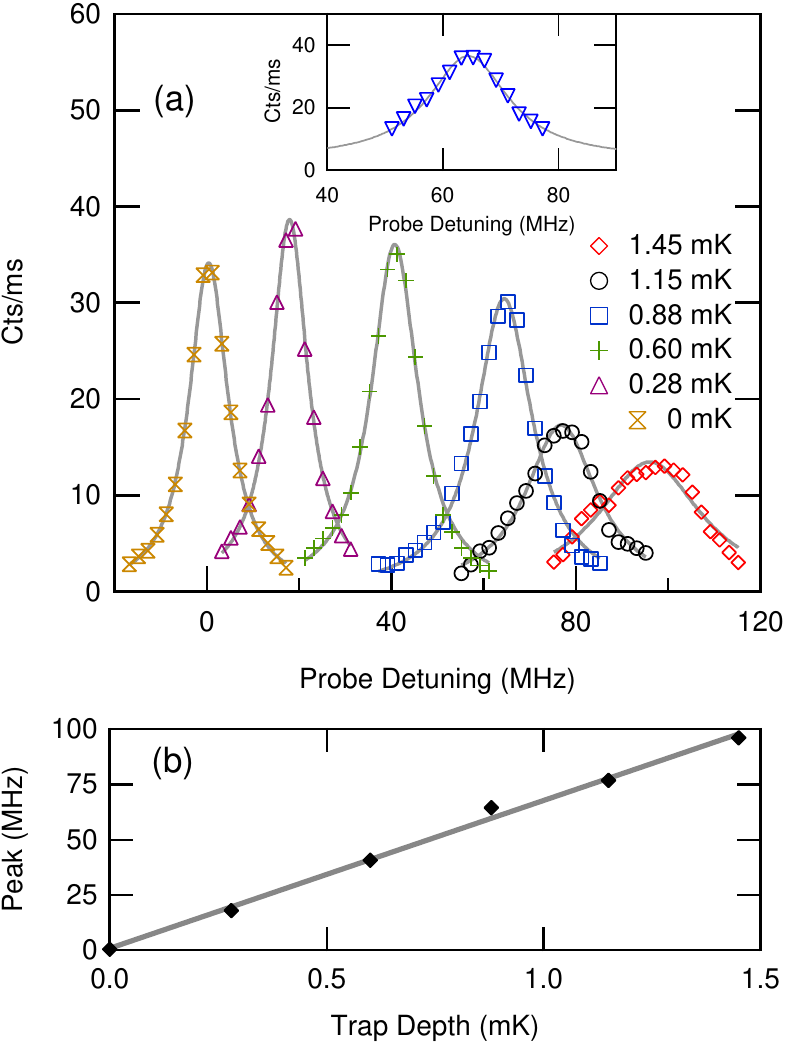}
\caption{\label{linearprobe} (Color online) (a) Spectrum measurements of single atoms in different trapping potentials using gated probing/cooling technique fit to the Lorentzian distribution. The on resonance probe intensity is $\simeq$$2I_{sat}$. The inset is a measurement with one-and-the-same atom in a 0.88 mK trap. The probe detuning starts from $+51$ MHz with an increment of 2~MHz, and the acquisition time at each detuning is 5~s. (b) Peak detuning versus trap depth; a least-squares fit to the data (solid line) indicates a slope of 67$\pm$2 MHz/mK.}
\end{figure}

Measurements of the AC-Stark shifted spectrum of single atoms in the FORT are shown in Fig.~\ref{linearprobe} for different trap depths. In these measurements, the trap beam and the probe beams are linearly polarized along the quantization axis ($y$-axis). The quantization axis is defined by zeroing the magnetic field to less than 10 mG using microwave spectroscopy, then a bias magnetic field of $\sim$500 mG is applied. In Fig.~\ref{linearprobe}(a), each data point is obtained by averaging the fluorescent signal from one and the same atom for 60~s, which gives a signal to noise ratio of 155 at the shifted resonance. For the 0~mK case, which corresponds to an untrapped atom, the trap beam is switched off for 1~$\mu$s during the probing period using the AOM, and the probing time is reduced to 500 ns. The spectrum for this measurement is fitted to a Lorentzian distribution centered at 0.4 MHz from the bare resonance, which is likely due to a small error of the probe laser frequency relative to the atomic transition. The inset of Fig.~\ref{linearprobe}(a) is a measurement of the AC-Stark shifted spectrum with one-and-the-same atom with 5~s acquisition time at each probe detuning. In general, the measurements are performed with one probe detuning at a time to maximize the signal to noise ratio.

\begin{figure}
\includegraphics[width=85mm]{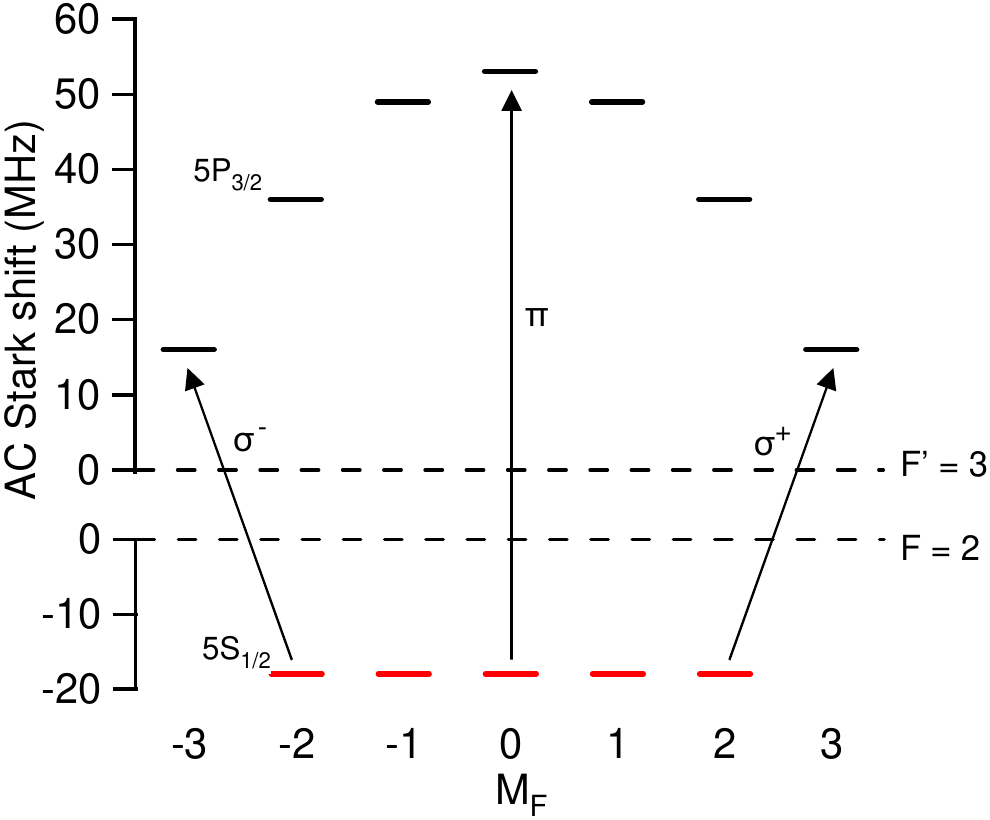}
\caption{\label{starkshiftcalculation} (Color online) The trapping potential used for this calculation is 0.88 mK. With trap beam linearly polarized along the quantization axis, the ground states are uniformly lowered by $h\times18$ MHz and the excited states are no longer degenerate.}
\end{figure}

The measured spectra are well fitted to Lorentzian distributions that are offset from the bare atomic resonance by an frequency that is proportional to the trap depth as shown in Fig.~\ref{linearprobe}(b). In order to compare the results in Fig.~\ref{linearprobe} to theory, we calculate the energy shifts for individual states within the hyperfine manifold, taking into account the multi-level structure of the atom and the dipole moments of the dipole-allowed transitions. The AC-Stark shift for the atoms in an optical dipole trap that is linearly polarized along the quantization axis with beam intensity $I$ and wavelength $\lambda$ is given by \citep{Safronova2011,Steck,Bransden}
\allowdisplaybreaks
\begin{align}
\Delta_{F,M_F} = &\frac{3{\pi}Ic^2}{2}\sum_{F',M_{F'}}^{}\frac{\Delta_{F'F}A_{F'F}}{\omega_{FF'}^3}\nonumber\\
&(2F+1)(2F'+1)(2J'+1)\nonumber\\
&\begin{pmatrix}
 F' & 1 & F\\
 M_{F'} & 0 &-M_F
\end{pmatrix}_{3j}^2
\begin{Bmatrix}
 J & J' & 1\\
 F' & F  &3/2
\end{Bmatrix}_{6j}^2,
\label{eq1}
\end{align}
with
\allowdisplaybreaks
\begin{align}
\label{eq2}
\Delta_{F'F} &= (\frac{1}{\omega_{FF'}+\omega}+\frac{1}{\omega_{FF'}-\omega}),\\
A_{F'F} &= \frac{2.02613\times10^{18}}{\lambda^3}\frac{2J+1}{2J'+1}d^2,
\label{eq3}
\end{align}
where $\omega_{FF'}$ is the transition angular frequency, $\omega$ is the trap laser angular frequency, $A_{F'F}$ is the transition rate, and $d$ is the electric dipole moment. Note that $\lambda$ is in {\AA} and $d$ is in atomic units where as Eqs. (1) and (2) are in SI units. The summation is carried over all dipole-allowed transitions from the state to be evaluated. Table \ref{tab:number} shows the transition, the corresponding wavelengths, and the electric dipole moments used for the calculation. Calculation of the $m_{F}$ state dependent AC-Stark shifts of the 5$S_{1/2}$, $F = 2$ ground states and the 5$P_{3/2}$, $F^{\prime} = 3$ excited states are shown in Fig.~\ref{starkshiftcalculation}. In the calculation, the trap laser wavelength is 1064 nm and the trap beam intensity is $5.7\times10^9$ W$\cdot$m$^{-2}$. The AC-Stark shift breaks the degeneracy of hyperfine sublevels of the 5$P_{3/2}$, $F^{\prime}$ = 3 states. On the other hand, the ground state shifts are uniform due to the geometry of the experiment, in which the trap beam is linearly polarized along the quantization axis.
The calculated AC-Stark shifts for the $\Delta M_F =0$ $\pi$ transitions are $+79$, $+74$, and $+60$ MHz/mK for  $\ket{F = 2, M_{F} = 0, \pm1, \pm2}$ $\rightarrow$ $\ket{F^{\prime} = 3, M_{F^{\prime}} = 0, \pm1, \pm2}$, respectively. The measured shift from Fig.~\ref{linearprobe}(b) is 67$\pm$2 MHz/mK, which is consistent with these calculations given the absence of optical pumping in the experiment and the uncertainty ($\sim$10\%) of the determination of the trap depths.

\begin{table}[h]
\begin{threeparttable}[b]
\caption{Transitions used for light shifts calculation, the corresponding wavelength in vacuum $\lambda$ and the electric dipole moment $d$.\label{tab:number}}
\begin{center}
\setlength{\tabcolsep}{16pt}
\begin{tabular}{|c r c|}
\hline
\centering Transition & \centering $\lambda$ (\AA)\tnote{a} & \centering $d$ ($ea_0$) \tabularnewline
\hline \hline
5$S_{1/2}$ $-$ 5$P_{1/2}$ & 7949.8 & $2.99$\tnote{a} \\
5$S_{1/2}$ $-$ 5$P_{3/2}$ & 7802.4 & $4.23$\tnote{a} \\
5$S_{1/2}$ $-$ 6$P_{1/2}$ & 4216.7 & $0.24$\tnote{a} \\
5$S_{1/2}$ $-$ 6$P_{3/2}$ & 4203.0 & $0.36$\tnote{a} \\
5$S_{1/2}$ $-$ 7$P_{1/2}$ & 3592.6 & $0.08$\tnote{a} \\
5$S_{1/2}$ $-$ 7$P_{3/2}$ & 7588.1 & $0.13$\tnote{a} \\
5$P_{3/2}$ $-$ 6$S_{1/2}$ & 13668.8 & $3.02$\tnote{b} \\
5$P_{3/2}$ $-$ 7$S_{1/2}$ & 7410.2 & $0.67$\tnote{b} \\
5$P_{3/2}$ $-$ 8$S_{1/2}$ & 6161.3 & $0.35$\tnote{b} \\
5$P_{3/2}$ $-$ 4$D_{3/2}$ & 15292.6 & $1.81$\tnote{b} \\
5$P_{3/2}$ $-$ 4$D_{5/2}$ & 15293.7 & $5.44$\tnote{b} \\
5$P_{3/2}$ $-$ 5$D_{3/2}$ & 7761.6 & $0.33$\tnote{b} \\
5$P_{3/2}$ $-$ 5$D_{5/2}$ & 7759.8 & $0.99$\tnote{b} \\
5$P_{3/2}$ $-$ 6$D_{3/2}$ & 6301.0 & $0.28$\tnote{a} \\
5$P_{3/2}$ $-$ 6$D_{5/2}$ & 6300.1 & $0.83$\tnote{a} \\
\hline
\end{tabular}
\begin{tablenotes}
\item[a] Ref. \cite{NIST}
\item[b] Ref. \cite{Clark2007}
\end{tablenotes}
\end{center}
\end{threeparttable}
\end{table}

\begin{figure}
\includegraphics[width=85mm]{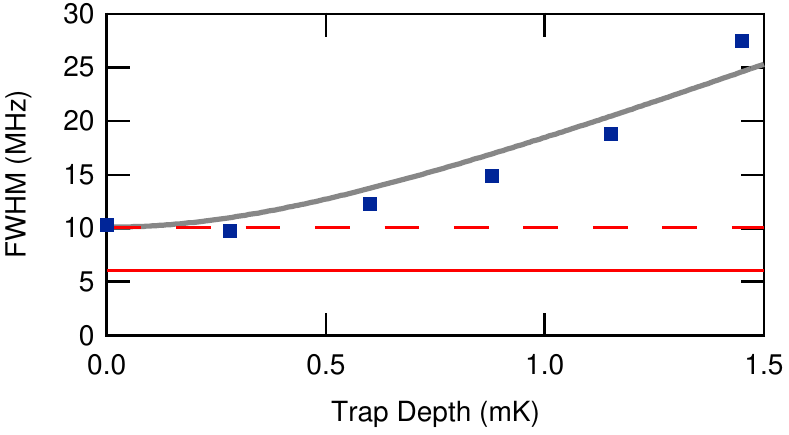}
\caption{\label{broadening} (Color online) FWHM of the single-atom spectrum in different trap depths as shown in Figure \ref{linearprobe}(a). The red solid line is the natural linewidth of the $^{87}$Rb 5$S_{1/2}$ $\leftrightarrow$ 5$P_{3/2}$ transition and the red dashed line is the power broadened linewidth with probe intensity $I$ = 2$I_{sat}$.}
\end{figure}

As shown in Fig.~\ref{broadening}, the transition linewidth broadens beyond the 6.1 MHz natural linewidth as the trapping potential increases. For the untrapped atom (0~mK), the measured linewidth is consistent with power-broadening, which for the probe intensity used in the experiment ($I$ = 2$I_{sat}$) yields a 10.3~MHz linewidth. The increase of linewidth broadening at higher trap depths can be attributed to a combination of averaging over the allowed $\pi$ transitions and thermal motion in the spatial varying light shifts of the potential.

\begin{figure}
\includegraphics[width=85mm]{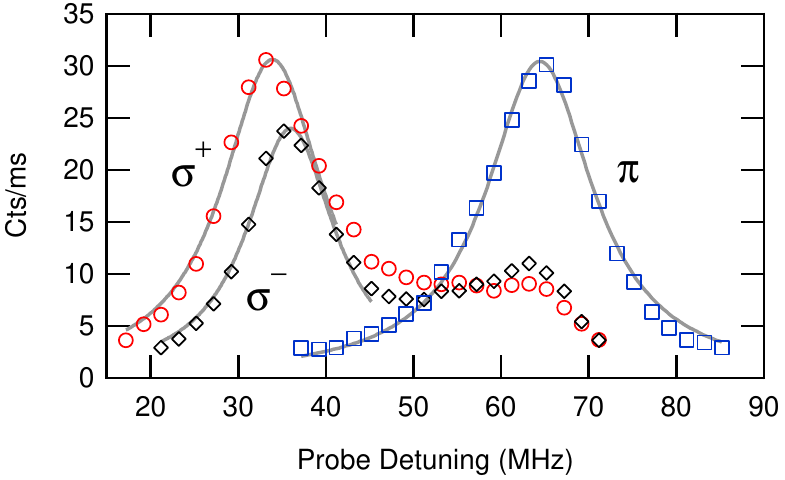}
\caption{\label{circularprobe} (Color online) The spectrum of single atom with circularly and linearly polarized probe. The trap depths are 0.88 mK for these measurements.}
\end{figure}

We have also measured the $M_F$ state dependent AC-Stark shifts by changing the polarization of the probe beams from linear to circular. The  spectra for the different polarizations are shown Fig.~\ref{circularprobe}. For the circular polarized probe beam case, the trap beam is linearly polarized along the probe beam propagation direction and the bias magnetic field direction ($z$-axis). The circularly polarized probe drives the atom to the stretched states, which shifts the measured peak to a lower frequency. Lorentzian fits to the central emission spectra indicate mean shifts of  ($\sigma^{+}$, $\sigma^{-}$, $\pi$) = (34, 36, 64) MHz with the same FWHM of 14 MHz. From the calculation, the $\ket{F = 2, M_{F} = \pm2}$ $\rightarrow$ $\ket{F^{\prime} = 3, M_{F^{\prime}} = \pm3}$ transitions are 34 MHz blue detuned from the bare resonant frequency. The 2 MHz difference between the results of $\sigma^{+}$ and $\sigma^{-}$ probe beams suggests that the trap laser is elliptically polarized and the trap beam propagation direction is not completely perpendicular to the quantization axis, which give rise to light shifts linearly dependent to $M_F$. The tails in the measured spectrum at higher frequency for the circularly polarized probes are likely due to either imperfect alignment between the bias magnetic field and the probe beam propagation direction or imperfect polarization of the probe and trap beams.

Finally, we mention that although the non-destructive method developed in this paper has been demonstrated using a single atom trap, it is possible to extend this method to resolved arrays of individual atoms. For this case, it will be necessary to detect the probe-induced emission on the camera to provide the necessary spatial resolution. Although the camera is not capable of gated operation, it is possible to separate the cooling and probe cycles spectrally by using the D1 and D2 transitions for the probe and cooling respectfully, and suitable narrow bandwidth filters to block the cooling emission from the camera.

In summary, we have demonstrated an efficient, non-destructive technique to characterize the AC-Stark shifts in single atom traps. By using short alternating probe and cooling pulses together with gated single-photon detection, we achieve high signal to noise ratio and long trap lifetimes. Using this method, we have measured the AC-Stark shifts of the $F=2 \rightarrow F'=3$ D2 transitions in ${}^{87}$Rb for different trap depths and probe polarizations. The measured spectra compare well to theoretical calculations. The development of non-destructive techniques to measure trap properties in situ will have important applications in scalable neutral atom quantum information experiments.

This work was supported by the National Science Foundation (Grant No. PHY-1107405).

\end{document}